\documentclass[a4paper,11pt]{article}
\pdfoutput=1 % if your are submitting a pdflatex (i.e. if you have
\usepackage{jinstpub} % for details on the use of the package, please
\usepackage{comment}

\usepackage[utf8]{inputenc}
\usepackage[T1]{fontenc}

\usepackage{url}
\newcommand{\cmmnt}[1]{\ignorespaces}

\title{\boldmath Performance and aging studies for the ALICE muon RPCs}

\author[a,1]{Luca Quaglia,\note{Corresponding author.}}
\author[a]{Antonio Bianchi,}
\author[a]{Alessandro Ferretti,}
\author[a]{Martino Gagliardi,}
\author[b]{Diego Stocco,}
\author[c]{Roberto Guida,}
\author[c]{Beatrice Mandelli,}
\author[d]{Laura Alvigini}

\affiliation[a]{INFN and University of Torino, via Pietro Giuria 1, Torino, Italy}
\affiliation[b]{SUBATECH, IMT Atlantique, Universit\'{e} de Nantes, CNRS-IN2P3, Nantes, France}
\affiliation[c]{EP-DT-FS Department, CERN, 1211 Geneva 23, Switzerland}
\affiliation[d]{IUSS and University of Pavia, Piazza della Vittoria 15, Pavia, Italy}

\emailAdd{luca.quaglia@unito.it}

\abstract{The ALICE muon trigger (MTR) system consists of 72 Resistive Plate Chamber (RPC) detectors arranged in two stations, each composed of two planes with 18 RPCs per plane. The detectors are operated in maxi-avalanche mode using a mixture of 89.7\% C$_{2}$H$_{2}$F$_{4}$, 10\% i-C$_{4}$H$_{10}$ and 0.3\% SF$_{6}$. A number of detector performance indicators, such as efficiency and dark current, have been monitored over time throughout the LHC Run2 (2015-18). While the efficiency showed very good stability, a steady increase in the absorbed dark current was observed. 

Since the end of 2018, the LHC has entered a phase of long shutdown, during which the ALICE experiment will be upgraded to cope with the next phase of data taking, expected in 2021. The MTR is undergoing a major upgrade of the front-end and readout electronics, and will change its functionalities, becoming a Muon Identifier \cite{i}. Only the replacement of the most irradiated RPCs is planned during the upgrade. It is therefore important to perform dedicated studies to gain further insights into the status of the detector. In particular, two RPCs were flushed with pure Ar gas for a prolonged period of time and a plasma was created by fully ionizing the gas. The output gas was analyzed using a Gas Chromatograph combined with a Mass Spectrometer and the possible presence of fluorinated compounds originating from the interaction of the plasma with the detector inner surfaces has been assessed using an Ion-Selective Electrode station.

This contribution will include a detailed review of the ALICE muon RPC performance at the LHC;
the procedure and results of the argon plasma test, described above, are also discussed.}


\proceeding{15$^{\text{th}}$ Workshop on Resistive Plate Chambers and Related Detectors\\
  when 10-14 February 2020\\
  where University of Rome Tor Vergata}

\begin{document}
\maketitle
\flushbottom

\section{Introduction}
\label{sec:intro}

A Large Ion Collider Experiment \cite{a} (ALICE) is one of the four main experiments located at the CERN Large Hadron Collider (LHC). It is specialized in the study of ultra-relativistic nucleus-nucleus collisions and it investigates the physics of strongly interacting matter at extreme energy densities, where the formation of the Quark Gluon Plasma \cite{b} (QGP) takes place. 

ALICE is equipped with a muon spectrometer covering the pseudorapidity interval -4<$\eta$<-2.5, whose primary aim is the measurement of muons from heavy flavours (hadrons containing a quark charm or beauty) and quarkonia (bound states $c\Bar{c}$ and $b\Bar{b}$).

The muon spectrometer is composed of a set of two absorbers (whose function is to reduce the flux of charged hadrons on the spectrometer), a tracking system made of ten detection planes, a dipole magnet and a set of four planes of single-gap Resistive Plate Chambers (RPCs) used to provide a trigger signal for the muon spectrometer. 

\section{The muon trigger system}
\label{sec:trig}
A detailed description of the muon spectrometer can be found in \cite{a}. The muon trigger system \cite{c} (MTR) is composed of a total of 72 single-gap RPCs, arranged in two stations (located at $\sim$16 and 17 m from the interaction point) with two detector planes in each (18 RPCs per plane). The gas gap is 2 mm thick, the electrodes are made out of low-resistivity (10$^{9}$-10$^{10}$ $\Omega \cdot$cm) bakelite and are 2 mm thick as well. The total active area per detection plane is $\sim$5.5x6.5 m$^{2}$.  
The detectors are operated in the so-called \textit{maxiavalanche} mode (average charge per hit of $\sim$100 pC/hit \cite{h,e,w}) with the following gas mixture: 89.7\% C$_{2}$H$_{2}$F$_{4}$, 10\% i-C$_{4}$H$_{10}$ and 0.3\% SF$_{6}$. The working voltage is set to values between 10 and 10.5 kV depending on the RPC. The detectors are read out by means of orthogonal copper strips and the signals are discriminated by the \textit{ADULT} \cite{d} front-end electronics, which has no amplification stage and threshold set to 7 mV.

%Each RPC of the muon trigger system is identified as follows: MT XY IN/OUT Z. MT stands for \textit{Muon Trigger}, X and Y are two numbers that can be either 1 or 2: X identifies the station and Y the planes. IN/OUT specifies if the RPC is inside or outside the LHC circumference and, lastly, Z is a number that goes from 1 to 9, from bottom to top, in order to distinguish each RPC in a given detection plane.

\begin{figure}[htbp]
\centering 
\includegraphics[width=.30\textwidth]{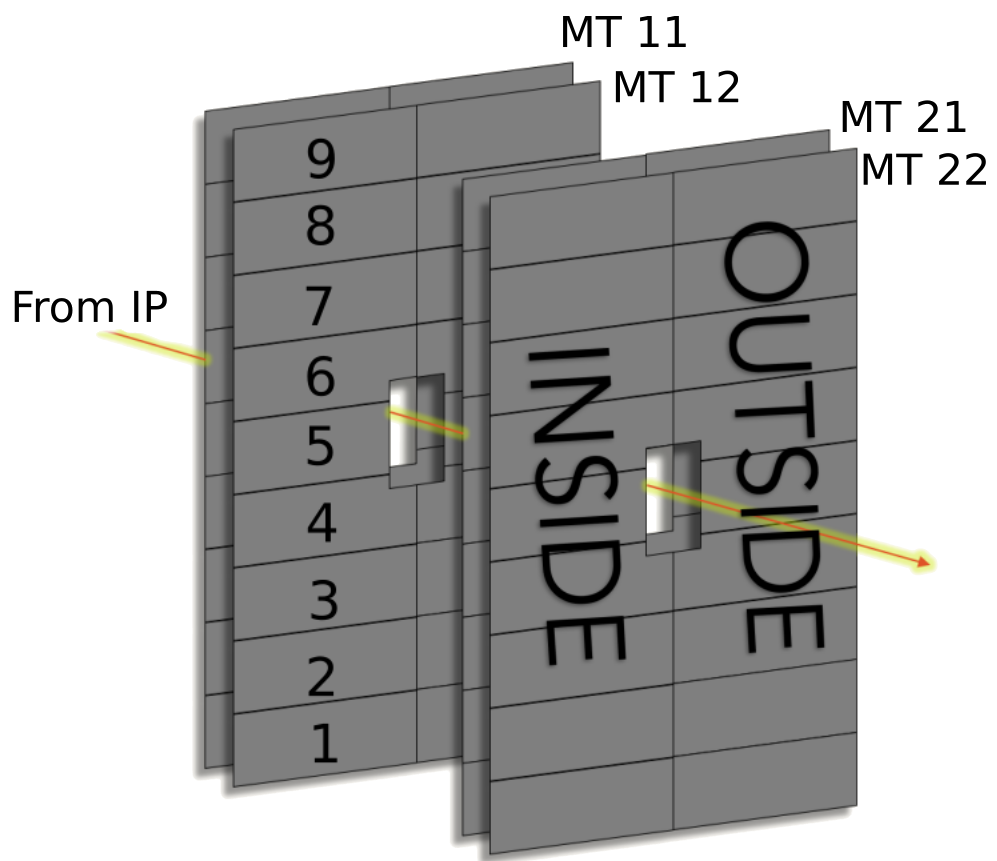}
\caption{\label{fig:0} Scheme of the RPCs in the ALICE muon trigger system with the adopted numbering scheme. \textit{Inside} and \textit{outside} refer to detection plane position with respect to the LHC circumference}
\end{figure}

%Starting from the LHC RUN 3 (2021-onward), the muon spectrometer will be operated in continuous readout mode (i.e. triggerless) and the muon trigger will become a muon identifier (MID) \cite{i}. The foreseen increase in collision rate will increase the maximum counting rate on the RPCs and it might also lead to a faster detector aging. For these reasons it is necessary to reduce the charge per hit liberated in the gas gap. This request will be satisfied thanks to a new front-end board called \textit{FEERIC} \cite{e} (Front-End Electronics Rapid Integrated Circuit) provided with an on-board pre-amplification stage. This will allow one to reduce the gain, hence the charge released in the gas gap.

\section{Performance of the ALICE MTR}
\label{sec:perf}

The ALICE experiment is operating since 2010. The data taking took place in two phases: the RUN 1 (2010-2013) and RUN 2 (2015-2018). During the operations, ALICE has taken data in different colliding systems: proton-proton (pp), proton-lead (p-Pb), lead-lead (Pb-Pb) and, for a very short period of time, xenon-xenon (Xe-Xe). The center of mass energies and integrated luminosities for the RUN 2 data taking period are reported in table \ref{tab:run}. The maximum instantaneous luminosity values are about 10$^{31}$ Hz/cm$^{2}$ for pp at $\sqrt{s}$=5.02 TeV, 5$\cdot$10$^{30}$ Hz/cm$^{2}$ for pp at $\sqrt{s}$=13 TeV, 1.5$\cdot$10$^{29}$ Hz/cm$^{2}$ for p-Pb and 10$^{27}$ Hz/cm$^{2}$ for Pb-Pb \cite{j}. 

\begin{table}[htbp]
\centering
	\caption{Running conditions during LHC RUN 2} 
	\label{tab:run}
	\smallskip
	\begin{tabular}{|c|c|c|c|} 
		\hline
		System & Year(s) & $\sqrt{s_{NN}}$ ($TeV$) & $L_{int}$ \\ [0.5ex] 
		\hline		
		pp & \begin{tabular}{@{}c@{}}2015,2017\\ 2015-2018 \end{tabular} & \begin{tabular}{@{}c@{}} 5.02 \\ 13\end{tabular} & \begin{tabular}{@{}c@{}} $\sim$ 1.5 $pb^{-1}$ \\ $\sim$ 36 $pb^{-1}$ \end{tabular} \\
		\hline
		p-Pb & 2016 & 5.02, 8.16 & $\sim$3 nb$^{-1}$, $\sim$ 25 nb$^{-1}$ \\
		\hline
		Pb-Pb &  2015,2018 & 5.02 & $\sim$800 $\mu$b$^{-1}$\\
		\hline
		Xe-Xe & 2017 & 5.44 & $\sim$0.3 $\mu$b$^{-1}$\\
		\hline
	\end{tabular}
\end{table}

\begin{figure}[htbp]
\centering % \begin{center}/\end{center} takes some additional vertical space
\includegraphics[width=.4\textwidth]{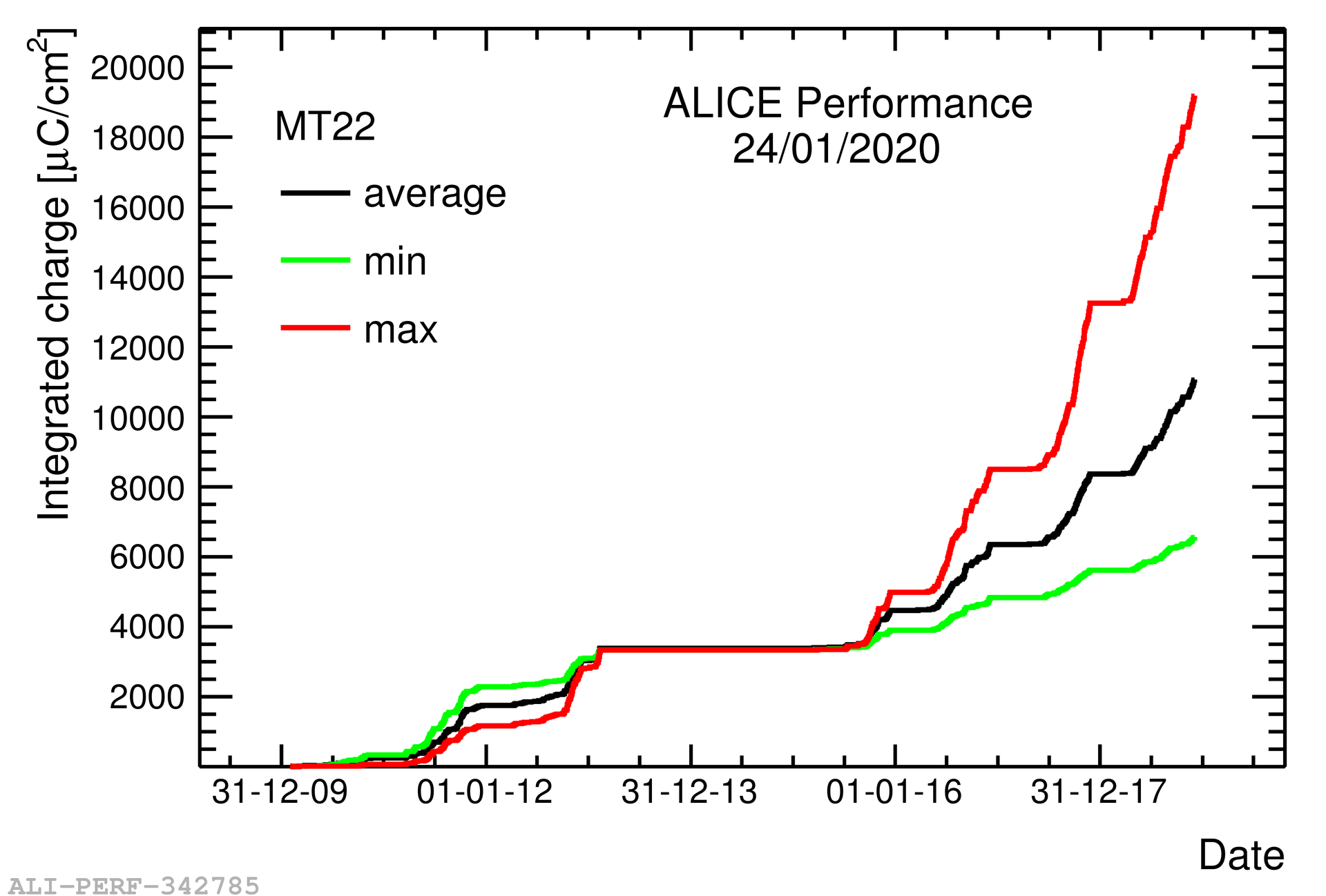}
\hspace{1cm}
\includegraphics[width=.4\textwidth]{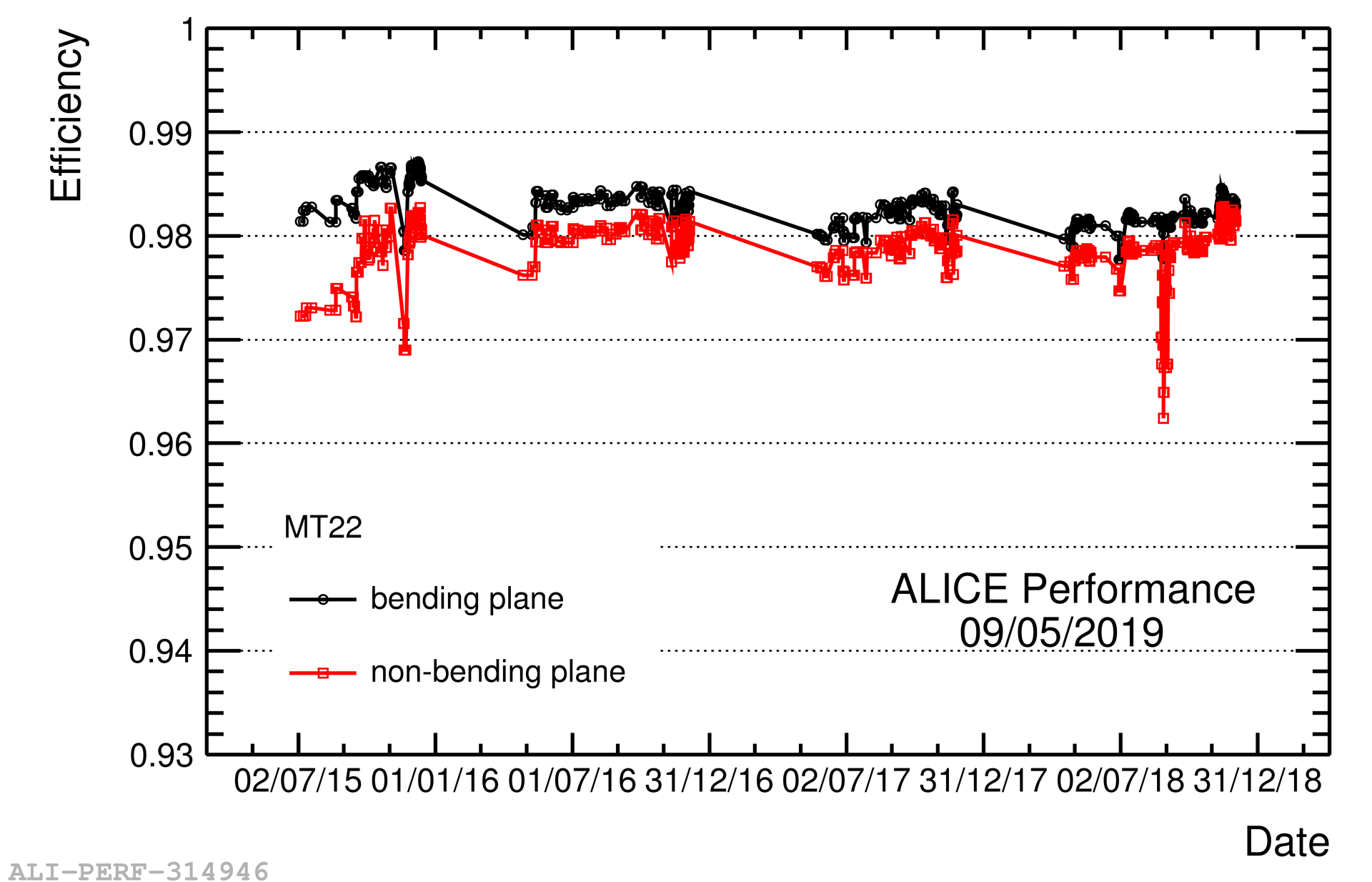}
\caption{\label{fig:1} Left panel: integrated charge trend for the MT 22 detection plane, right panel: efficiency trend for the MT 22 detection plane}
\end{figure}

The charge integrated by the RPCs has been monitored since the beginning of operations, using continuous current measurements after dark current subtraction. The latter is defined as the current absorbed by the detectors when they are not irradiated (i.e. due to intrinsic noise and cosmic rays) and is estimated through dedicated cosmic rays runs. The left panel of figure \ref{fig:1} shows the trend of integrated charge for the MT 22 detection plane (see figure \ref{fig:0}) as a function of time. Three curves are shown: one for the average integrated charge, and two for the RPCs that had accumulated the highest and the lowest amount of charge by the end of RUN 2. The MT 22 plane was chosen as an example because it was the one which, on average, accumulated the greatest amount of charge ($\sim$11 mC/cm$^2$). Aging tests \cite{h} certified RPC operation with the ALICE gas mixture up to $\sim$ 50 mC/cm$^{2}$ of integrated charge.

The RPC efficiency is constantly monitored, to ensure that the detectors are working as expected and is an input for efficiency corrections in data analysis. In the right panel of figure \ref{fig:1} the trend of the average efficiency for the MT 22 detection plane is shown for the whole Run 2. The curve in red refers to the non-bending plane, while the black one to the bending plane. Bending and non-bending refer to the strip orientation relative to the dipole magnetic field: the strips parallel (perpendicular) to the magnetic field form the bending (non-bending) plane. The Efficiency showed satisfactory results for the whole detection system, being typically > 96\% and stable over time. The small fluctuations are mainly due to local issues, such as noise in the front-end electronics. The MT 22 detection plane was chosen as an example but the behavior of the other planes is similar. 

The detector availability for data taking during RUN 2 was > 95\% (the missing 5\% also includes those runs in which the RPCs were kept OFF due to unavailability of other detectors used for muon physics).

\begin{figure}[htbp]
\centering % \begin{center}/\end{center} takes some additional vertical space
\includegraphics[width=.35\textwidth]{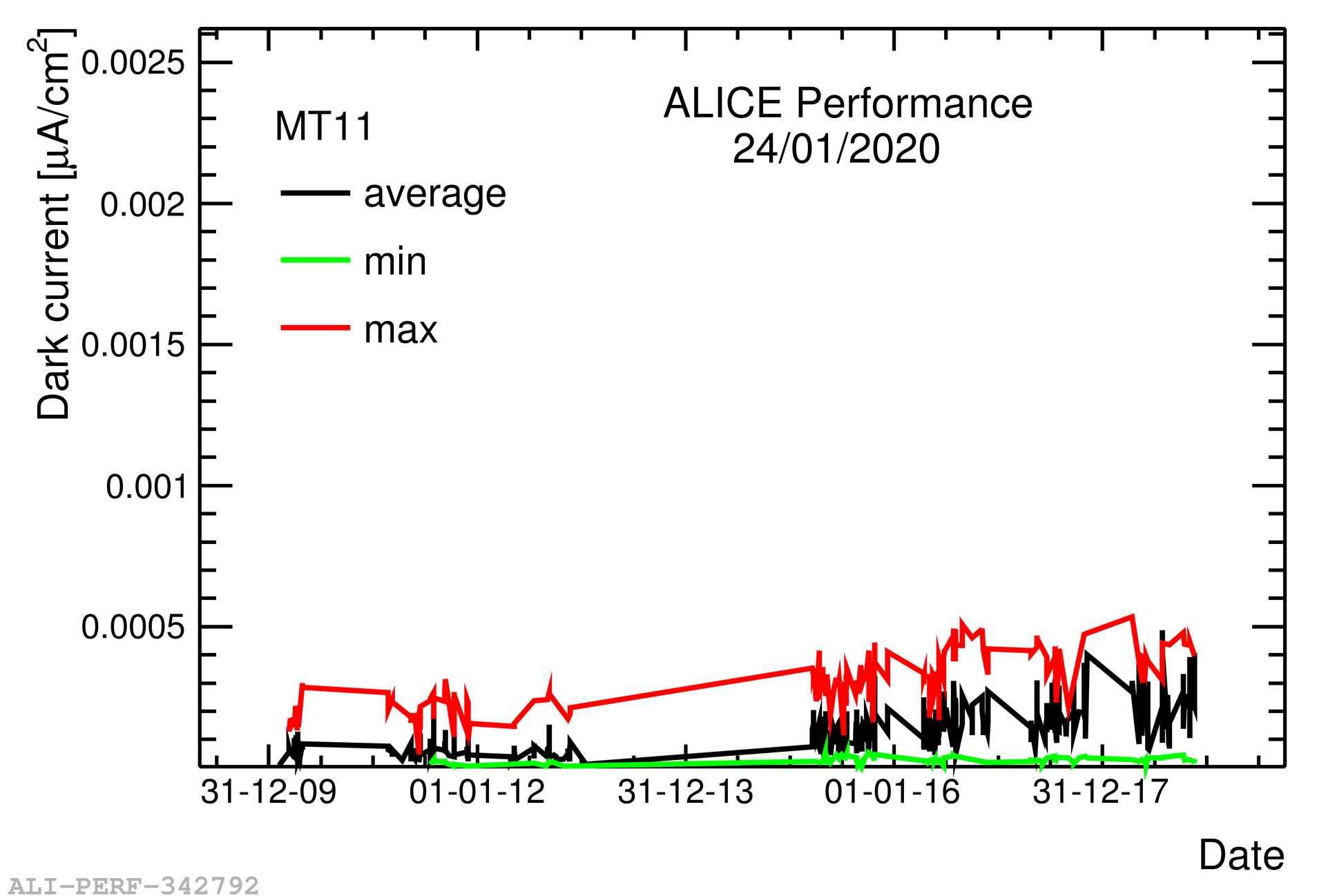}
%\qquad
\includegraphics[width=.35\textwidth]{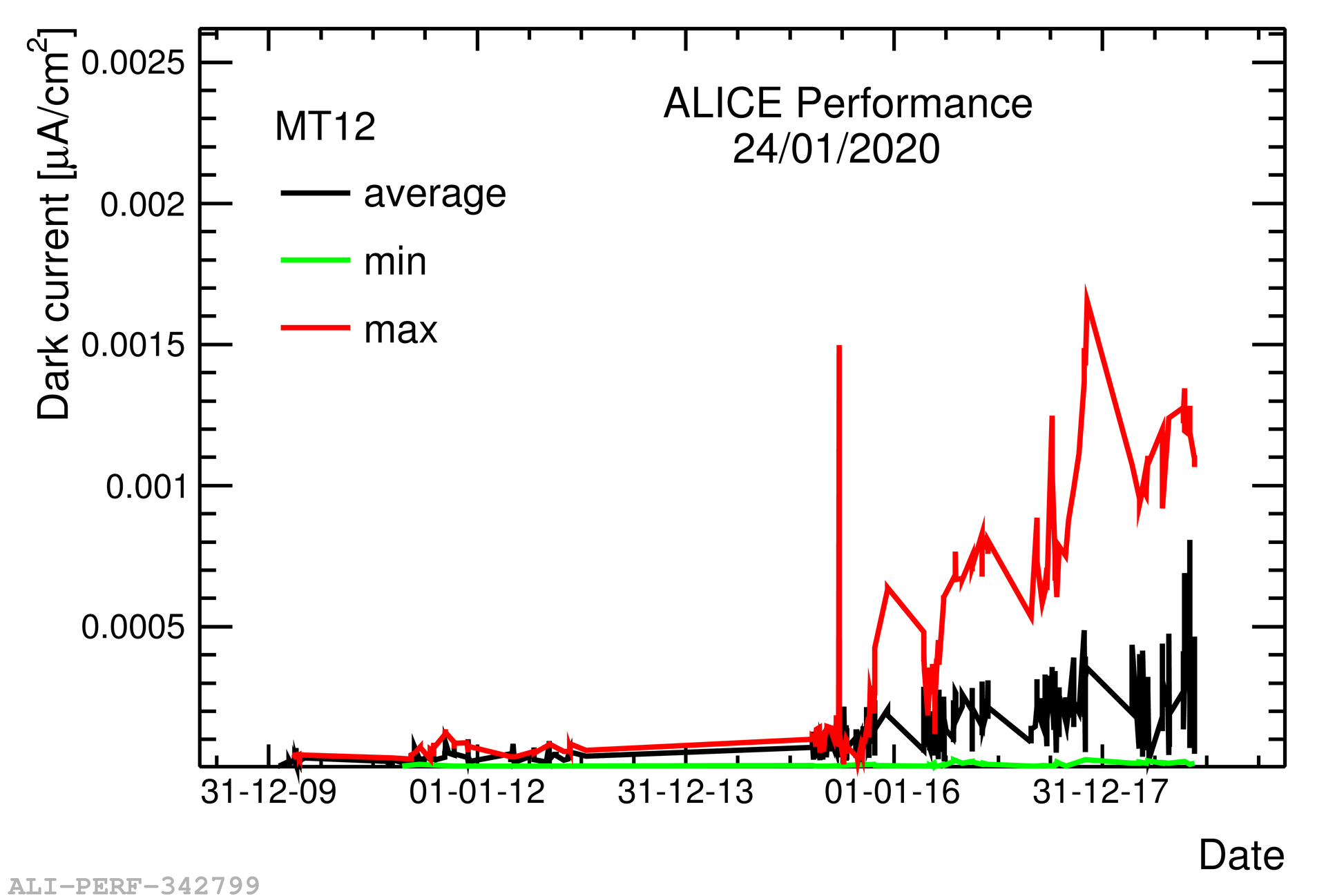}
\includegraphics[width=.35\textwidth]{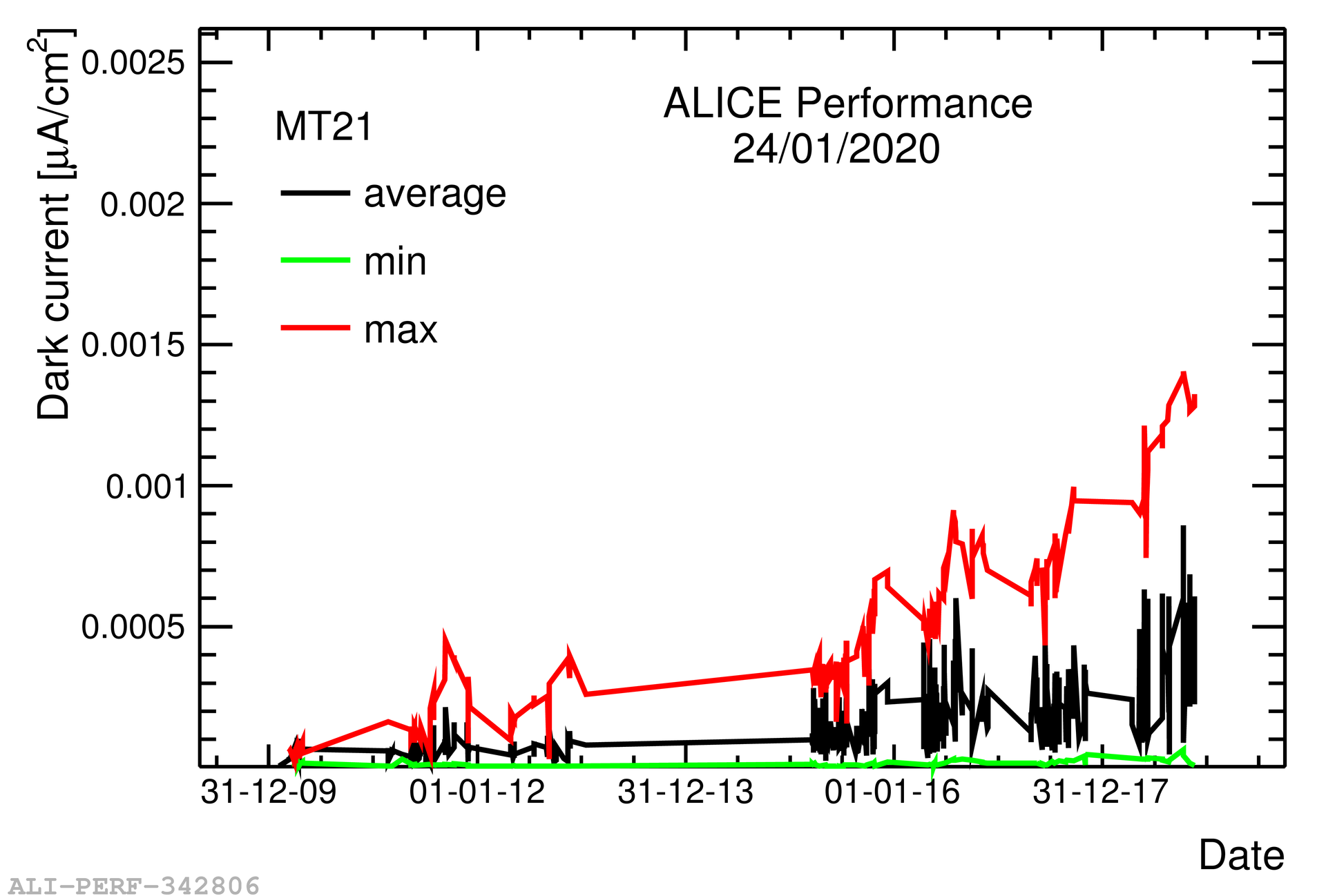}
%\qquad
\includegraphics[width=.35\textwidth]{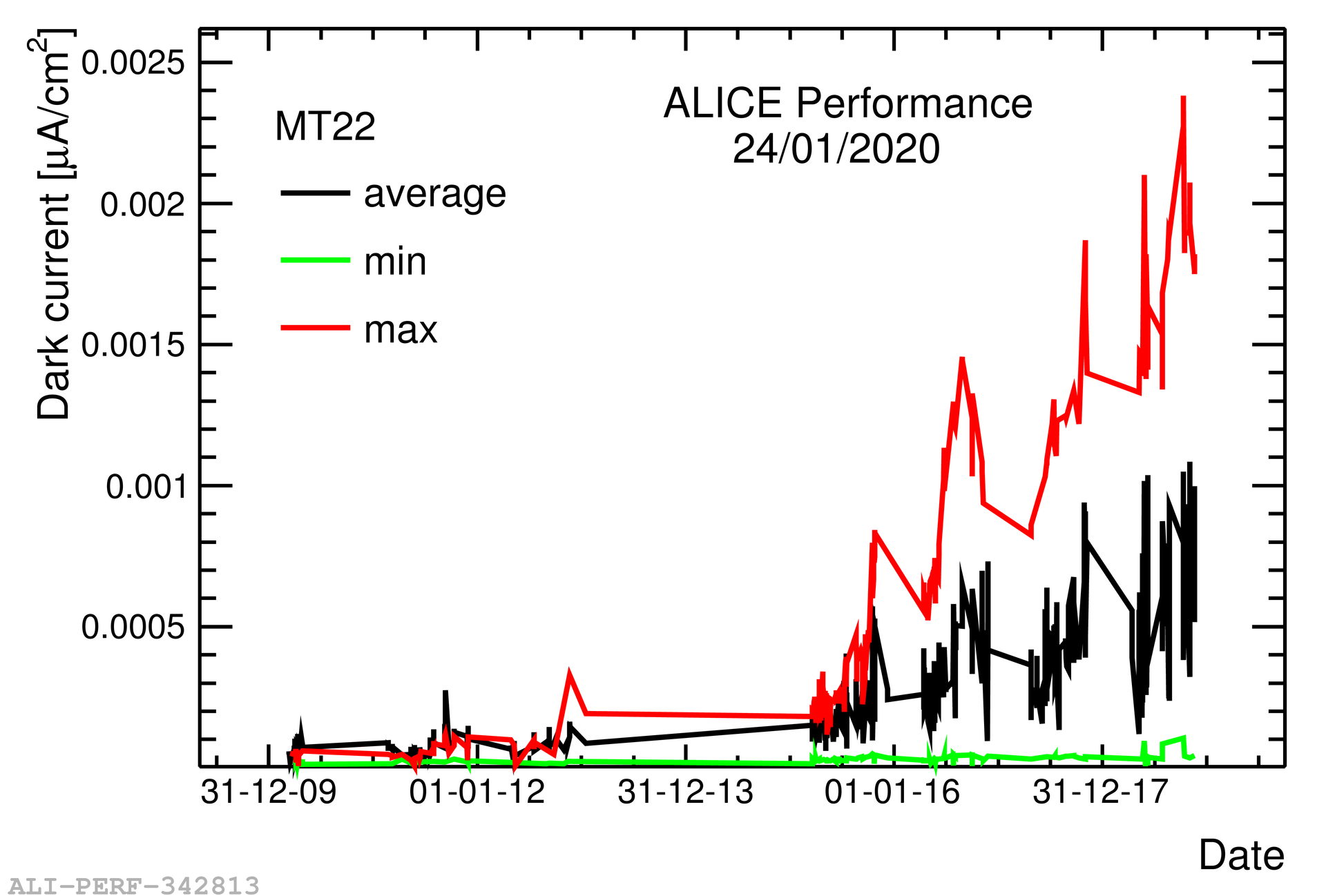}
% "\includegraphics" from the "graphicx" permits to crop (trim+clip)
% and rotate (angle) and image (and much more)
\caption{\label{fig:3} Dark current trend for the ALICE muon RPCs.}
\end{figure}

An increase in the average values of dark current can be observed for all four detection planes, especially during RUN 2, as it can be seen in figure \ref{fig:3}. This is the only sign of potential aging shown by the system since, as stated earlier, the efficiency does not deteriorate over time. The increase in absorbed dark current was observed on $\sim$50\% of the detectors and the causes are under investigation.

\section{Test with Ar plasma}
\label{sec:aging}

A possible explanation for the observed dark current increase is the deposition of fluorinated compounds, mainly hydrofluoric acid (HF), on the inner surfaces of the detectors. These compounds are created by the recombination of F$^{-}$ ions (originated by the breakage of C$_{2}$H$_{2}$F$_{4}$ molecules following their interaction with radiation) with hydrogen, originating mainly from the water vapour in the mixture. This compound may, at least partially, deposit on the inner surfaces of the detectors, leading to a degradation of these surfaces and giving rise to the increase in absorbed dark current, e.g. by chemically attacking the electrodes surface creating local spikes or by forming a thin conductive layer \cite{z}.

%breakage of the chemical bonds in the gas molecules due to their interaction with radiation, in a phenomenon known as \textit{gas radiolysis}. Since the main component of the gas mixture (89.7\%) is C$_{2}$H$_{2}$F$_{4}$, many F$^{-}$ ions are produced and they can recombine with hydrogen from the water vapour in the mixture, creating HF. This compound is, in part, swept away from the gas flow but it can, in principle, deposit on the inner surfaces of the detectors, creating a thin conductive layer and giving rise to the increase in absorbed dark current.

Under this hypothesis, an argon plasma test \cite{k} was executed on two of the MTR RPCs (active area of 292x77 cm$^{2}$ each): these detectors were flushed with pure argon and a plasma was created by ionizing it. A few other detectors (9) were flushed with argon but without ionizing it and the others were left untouched. The free charges flowing in the plasma, as well as the photons originating from ion-electron recombinations, may be sufficiently energetic to detach the fluorinated compounds from the inner surfaces of the detectors and the gas flow would take them out. 

\begin{figure}[htbp]
\centering % \begin{center}/\end{center} takes some additional vertical space
\includegraphics[width=.5\textwidth]{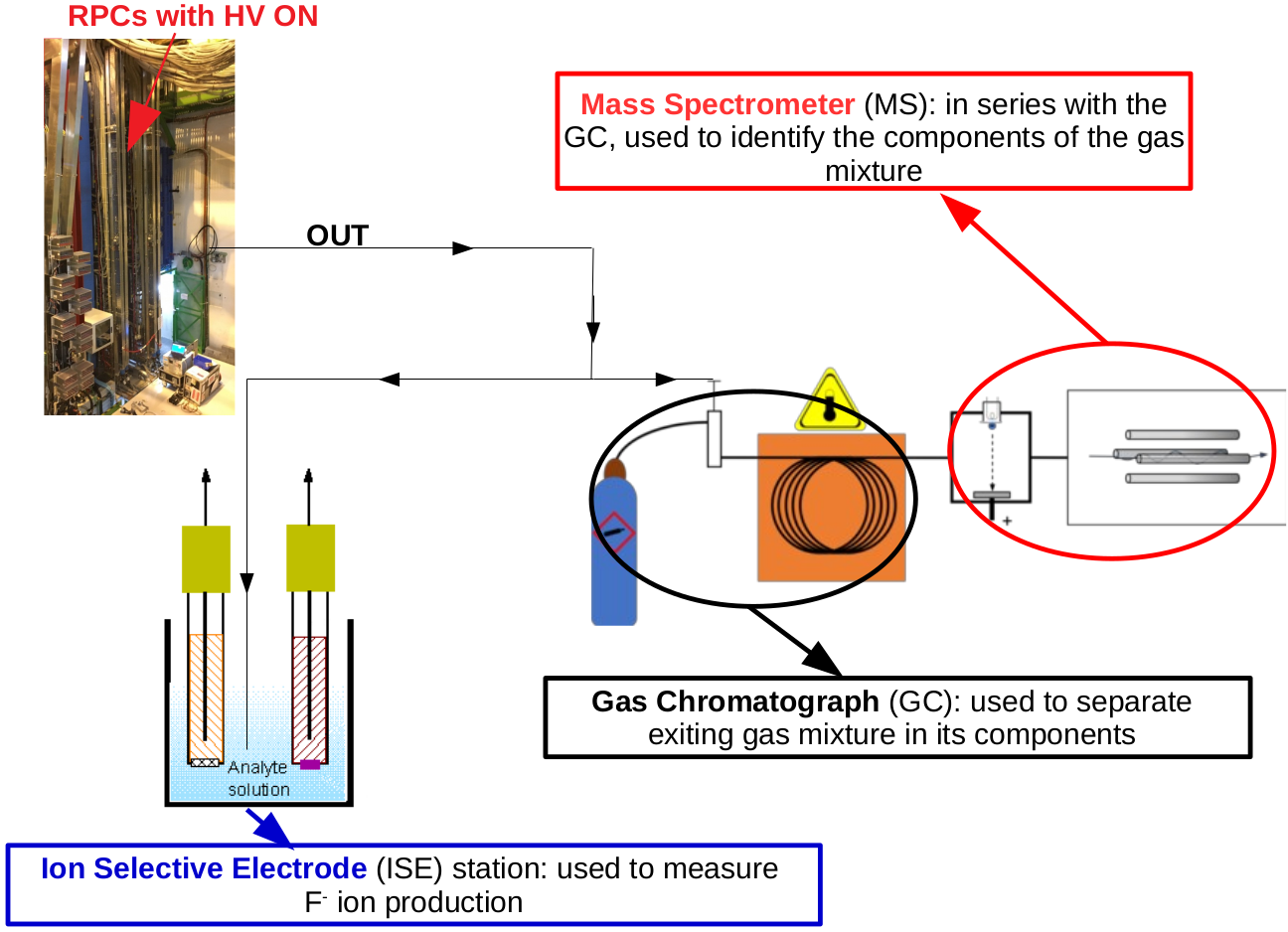}
\caption{\label{fig:4} Scheme of the experimental apparatus for the Ar plasma test.}
\end{figure}

To confirm this assumption, analyses of the exiting gas mixture were performed using a Gas Chromatograph/Mass Spectrometer (GC/MS) combination, in order to identify the presence of compounds potentially produced by the interaction of the plasma with the detector, and an Ion Selective Electrode (ISE) station to identify the presence of F$^{-}$ ions in the exiting gas mixture. The GC performs the separation of a given gas mixture in its components, thanks to the fact that different gases are trapped for different times in the GC elements. Each gas exits the GC after a specific time (retention time) and enters the mass spectrometer (MS) in order to be identified. Each component of the gas mixture produces a peak in a chromatogram (the result of a GC analysis) and the area underneath it (measured in $\mu$V$\cdot$s) is proportional to the concentration of that particular element. Figure \ref{fig:4} shows a scheme of the experimental apparatus.

A comparison between the dark current absorbed with the standard ALICE gas mixture at the working point, before and after the Ar plasma test, was also carried out in order to assess whether this procedure had any effect on it.

\subsection{F$^{-}$ ions production}
\label{sub:fluoride}

\begin{figure}[htbp]
\centering % \begin{center}/\end{center} takes some additional vertical space
\includegraphics[width=.43\textwidth]{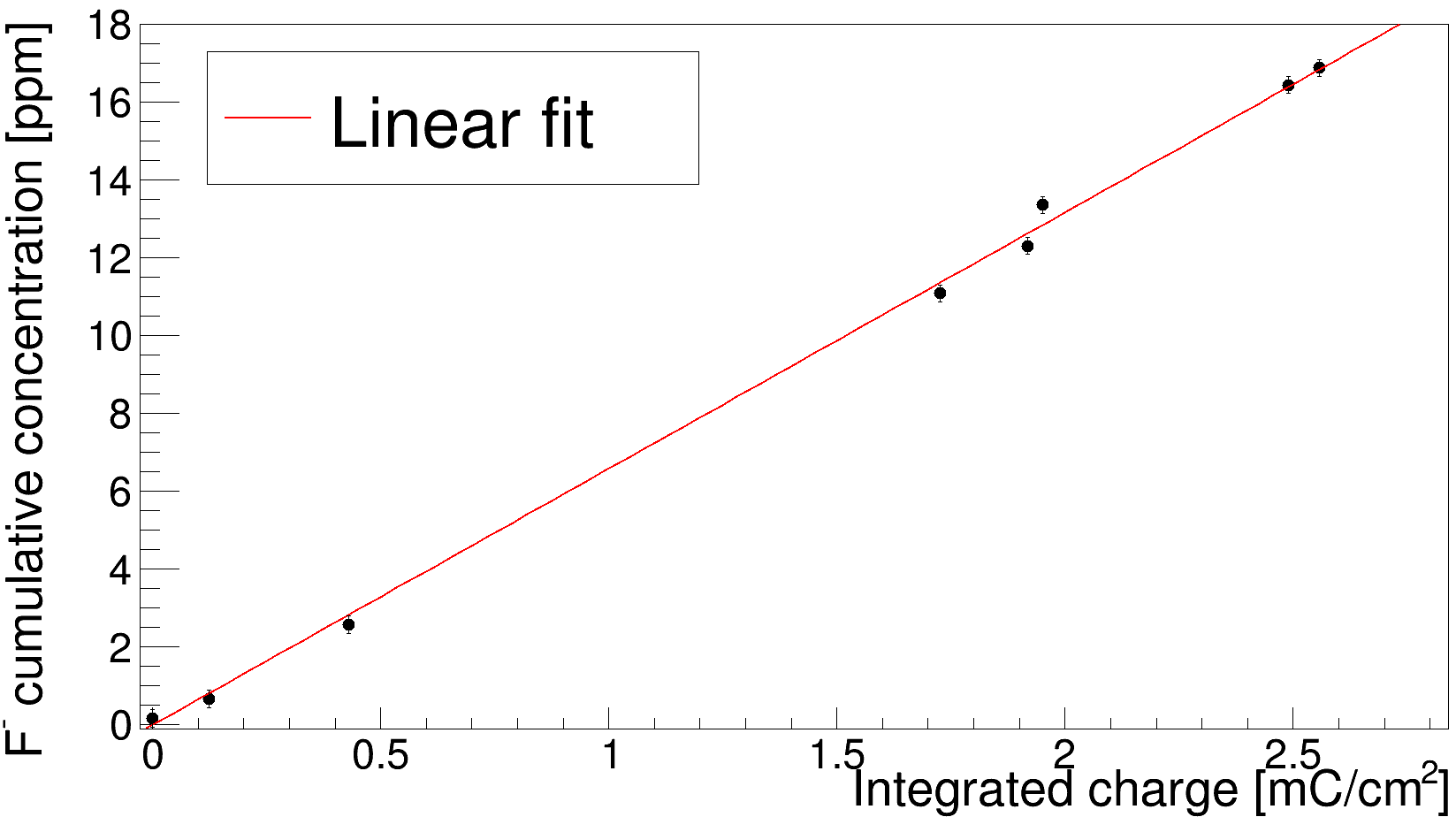}
\caption{\label{fig:fluoro} Trend of cumulative F$^{-}$ ions concentration as a function of integrated charge.}
\end{figure}

The ISE produces a voltage signal proportional to the F$^{-}$ ions concentration which is then converted into a concentration value (measured in ppm) via a calibration curve. During the Ar plasma test, integrated measurements were carried out, meaning that the exiting gas was bubbled for a prolonged period of time (a few hours) in 33 ml of distilled water and a measurement of the F$^{-}$ ions accumulated in that period of time was done. In figure \ref{fig:fluoro}, the cumulative concentration of fluoride ions, gathered from different measurements, is plotted as a function of the charge integrated during the test.  

The F$^{-}$ concentration was measured before the beginning of the Ar plasma test and found to be zero, while during the test the concentration was found to increase linearly with the integrated charge. The most natural explanation is the production of fluorinated compounds, due to the interaction of the Ar plasma with the inner surfaces of the detectors. The production rate of F$^{-}$ seems to be constant during the whole test, indicating that the detachment process was still ongoing after an integrated charge of $\sim$2.5 mC/cm$^{2}$.

\subsection{Resistivity studies}
\label{sub:res}

When operated with pure Ar, the characteristic current-voltage (I-V) curve for an RPC has the following features: up to $\sim$2000 V the Ar is not ionized and no current is circulating in the gas gap. The gas ionizes rather quickly above 2000 V and, when it reaches full ionization, the I-V curve follows Ohm's law. The gas behaves like a short circuit between the two bakelite electrodes, which can be considered as two resistors in series. From a linear fit to the I-V curve it is possible to extract the total resistance of the two bakelite electrodes and calculate the mean resistivity of the detector. 

\begin{figure}[htbp]
\centering % \begin{center}/\end{center} takes some additional vertical space
\includegraphics[width=.4\textwidth]{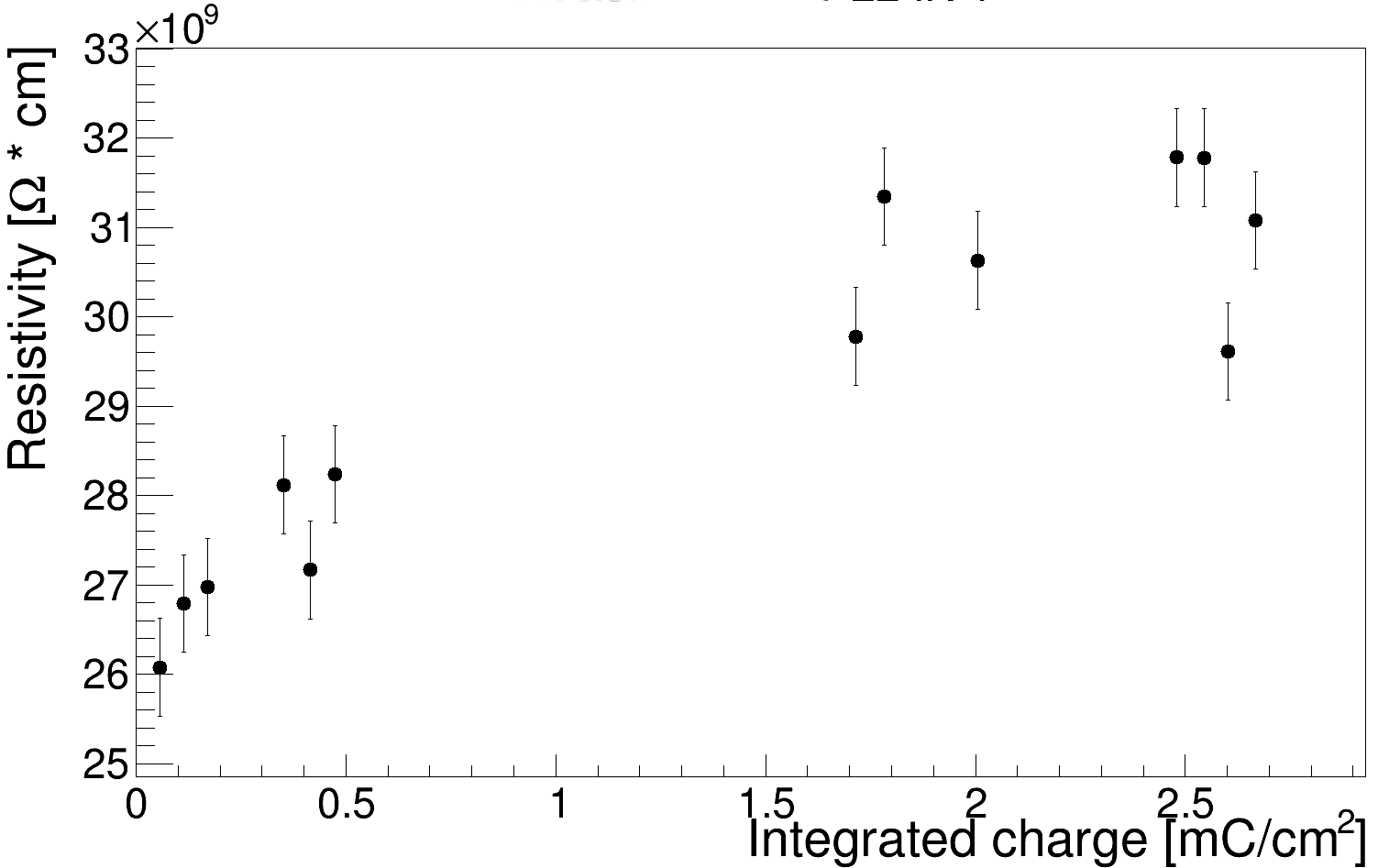}
\hspace{1 cm}
\includegraphics[width=.4\textwidth]{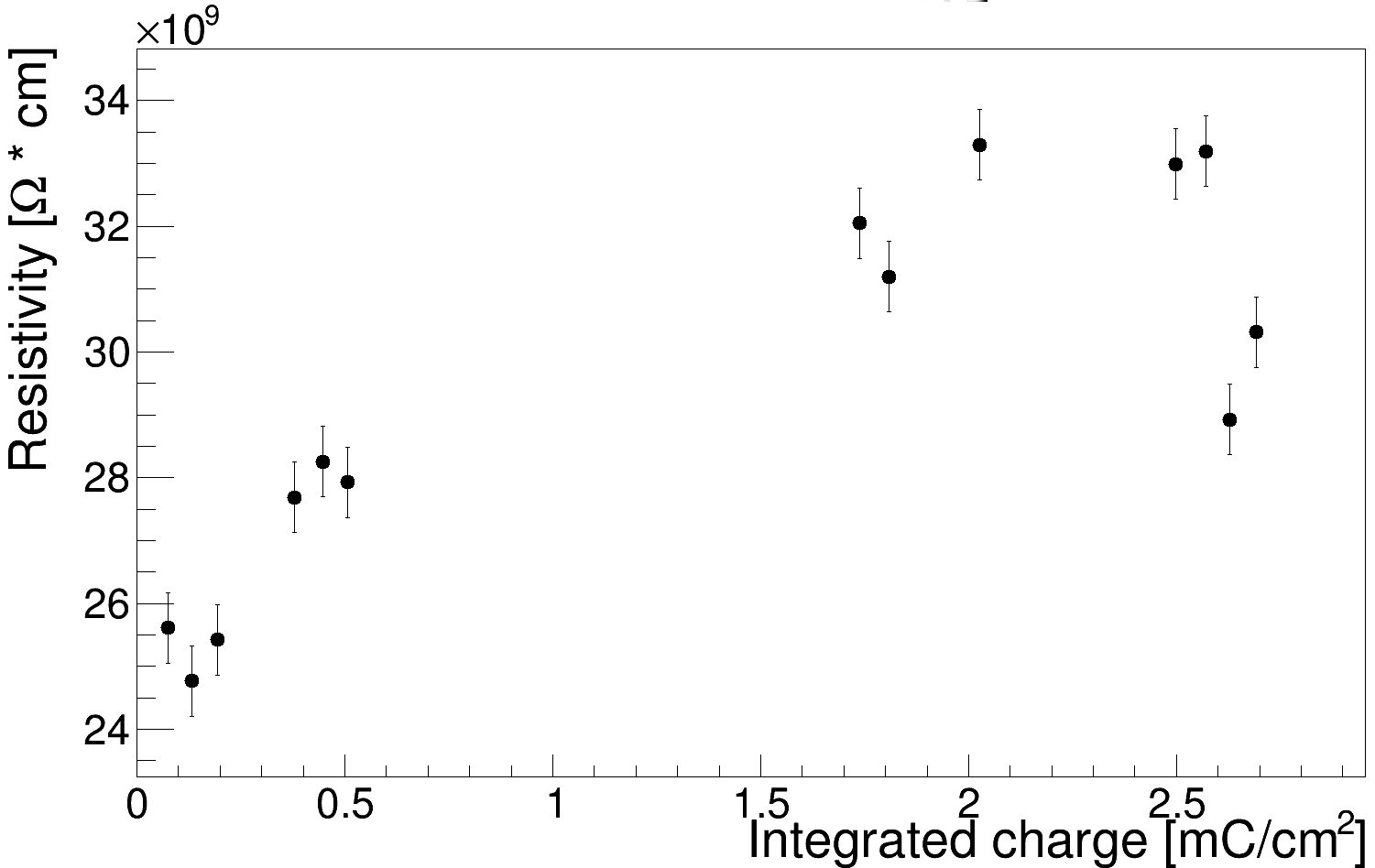}
\caption{\label{fig:res} Left: resistivity trend for the two chambers under test}
\end{figure}

In figure \ref{fig:res}, the resistivity of the two detectors under test is shown as a function of the charge accumulated during the Ar plasma test. The resistivity, for both chambers, shows hints of an increasing trend with the integrated charge. This might be related to a drying effect that the plasma is having either on the bakelite or on the linseed oil coating, as it is suggested in \cite{g}.

\subsection{CO$_{2}$ production}
\label{sub:co2}

Besides Ar and water, already present at the input of the detectors, another compound was identified by the GC/MS analyses, when the detectors were switched on and the plasma was created. This was identified as CO$_{2}$.

\begin{figure}[htbp]
\centering % \begin{center}/\end{center} takes some additional vertical space
\includegraphics[width=.42\textwidth]{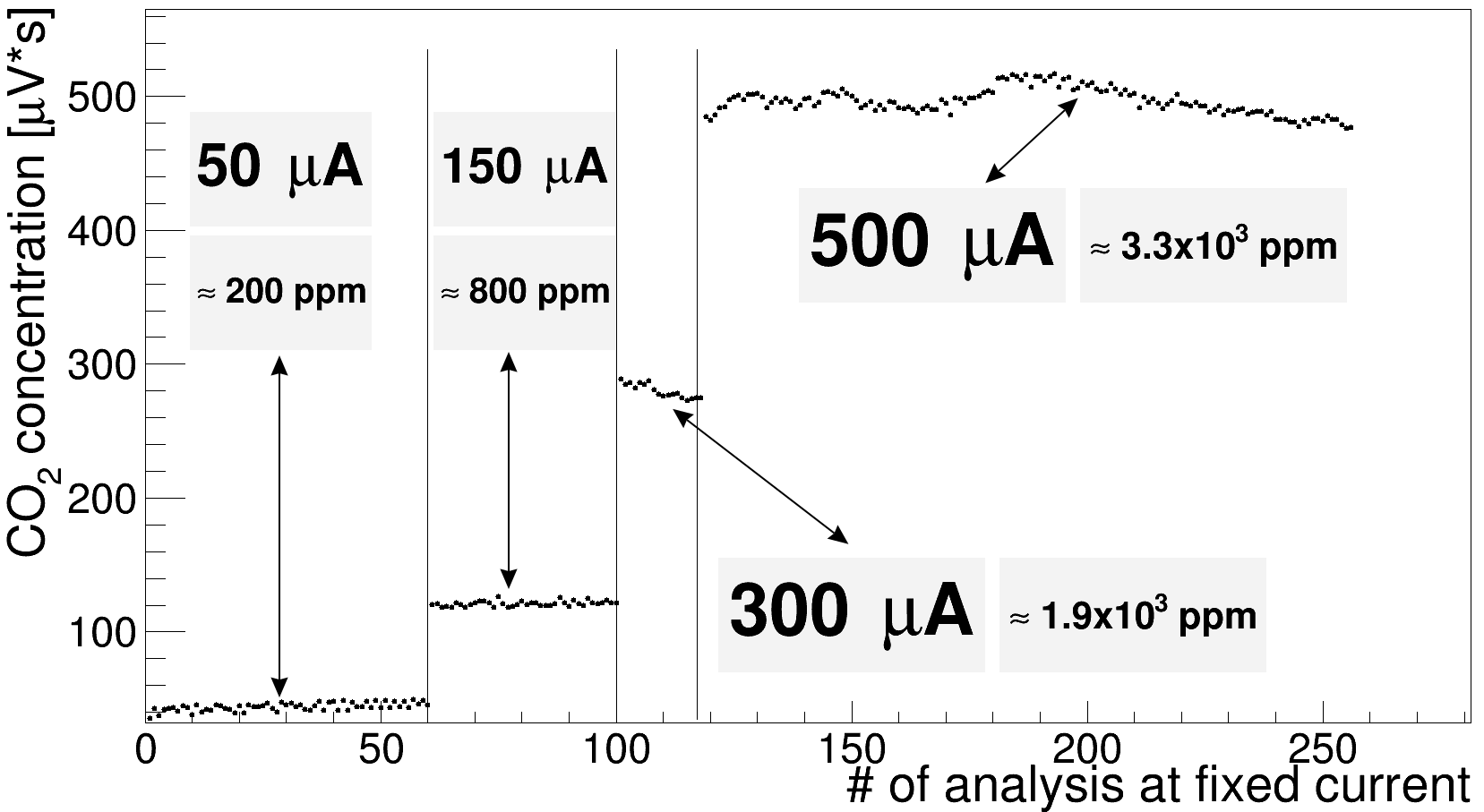}
\hspace{1 cm}
\includegraphics[width=.42\textwidth]{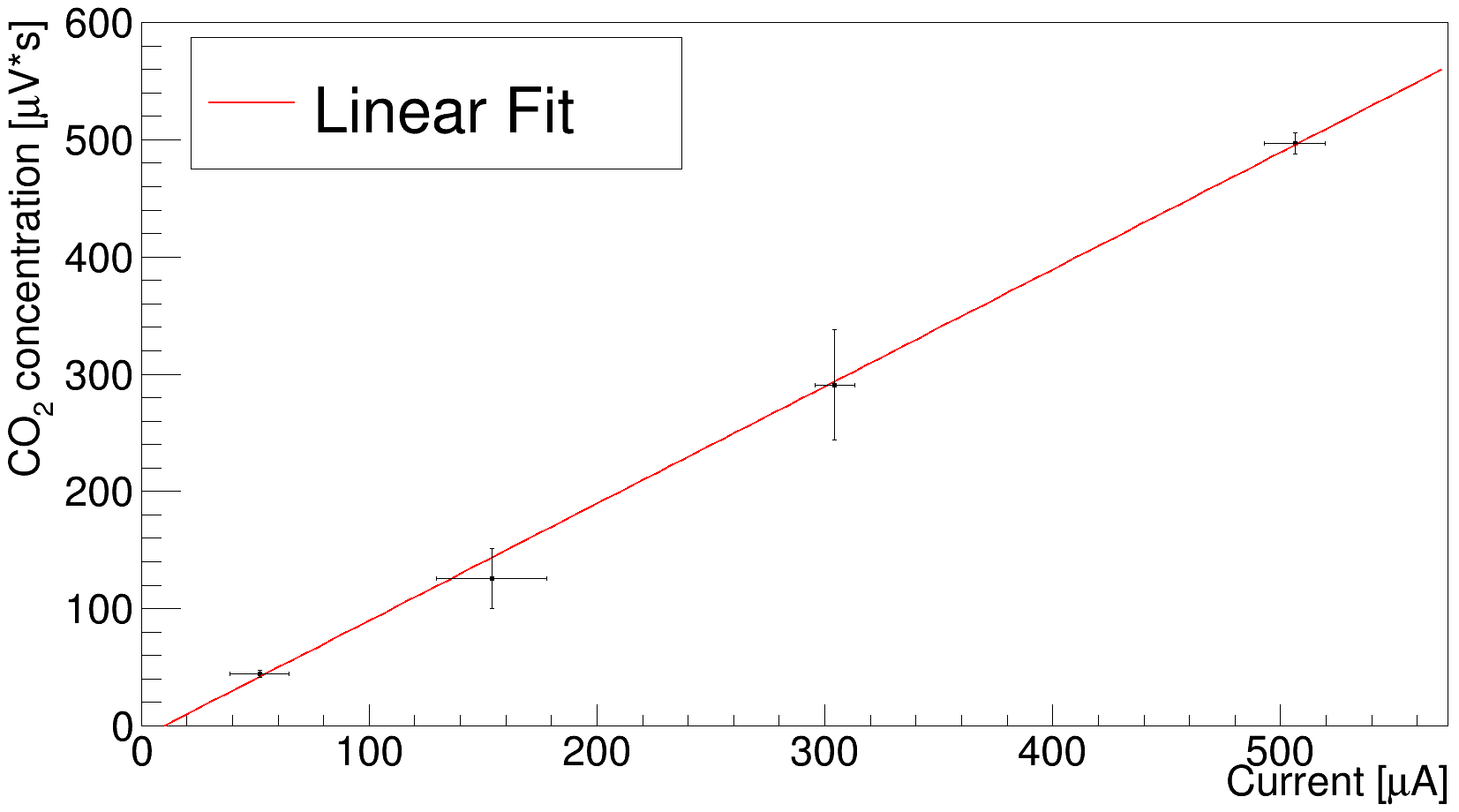}
\caption{\label{fig:lin} Correlation between CO$_{2}$ concentration and circulating current. The values in ppm for the CO$_{2}$ concentration have been obtained by means of a calibration curve.}
\end{figure}

The CO$_{2}$ concentration was correlated with the circulating current, as shown in figure \ref{fig:lin}: measurements have been taken at different values of current and the CO$_{2}$ concentration was calculated for each analysis. In the left portion of figure \ref{fig:lin} all the results are shown, while in the right portion the average values of concentration are plotted as a function of the average values of applied current. The linear fit clearly shows a correlation between the two quantities.  

\subsection{Dark current comparison}
\label{sub:dark_comp}

The dark current with the standard gas mixture at the working point was measured before and after the tests. The observed small variations are compatible with those observed for the RPCs which were not treated.

\section{Conclusions}
\label{sec:concl}

The RPCs of the ALICE muon trigger system have shown satisfactory results during the LHC RUN 2: efficiency was > 96\% and stable over time, detector availability was over 95\%; some of the detectors have accumulated an amount of charge close to their certified lifetime and may have to be replaced before LHC RUN 3 \cite{i}.

An increase in the absorbed dark current was observed. In order to gain further insights on such an increase, an argon plasma test was carried out and the following observations have been made: (i) F$^{-}$ ions were present in the exiting gas mixture and their concentration was correlated with the charge accumulated in the test, (ii) hints of an increasing trend of the RPCs resistivity, (iii) the production of CO$_{2}$ and its correlation with the circulating current, (iv) a comparison between the dark current before and after the test showed no significant difference.

\end{document}